\documentclass[showpacs,twocolumn,preprintnumbers,amsmath,amssymb,prl,epsf,superscriptaddress]{revtex4-1}
\usepackage{graphicx}
\usepackage{nicefrac}
\usepackage[squaren]{SIunits}
\usepackage{color}
\begin{document}
\title{Low-noise macroscopic twin beams}
\author{Timur Sh.~Iskhakov}
\affiliation{Max Planck Institute for the Science of Light,
G\"unther-Scharowsky-Stra\ss{}e 1/Bau 24, 91058 Erlangen, Germany}
\author{Vladyslav~C.~Usenko}
\affiliation{Department of Optics, Palack\'y University, 17. Listopadu 50, 772 00 Olomouc, Czech Republic}
\affiliation{Bogolyubov Institute for Theoretical Physics, Metrolohichna str. 14-b, 03680 Kiev, Ukraine}
\author{Radim Filip}
\affiliation{Department of Optics, Palack\'y University, 17. Listopadu 50, 772 00 Olomouc, Czech Republic}
\author{Maria~V.~Chekhova}
\affiliation{Max Planck Institute for the Science of Light, G\"unther-Scharowsky-Stra\ss{}e 1/Bau 24, 91058
Erlangen, Germany} \affiliation{Department of Physics, M.V.Lomonosov Moscow State University, \\ Leninskie Gory,
119991 Moscow, Russia}\affiliation{University of Erlangen-N\"urnberg, Staudtstrasse 7/B2, 91058 Erlangen, Germany}
\author{Gerd Leuchs}
\affiliation{Max Planck Institute for the Science of Light, G\"unther-Scharowsky-Stra\ss{}e 1/Bau 24, 91058
Erlangen, Germany} \affiliation{University of Erlangen-N\"urnberg, Staudtstrasse 7/B2, 91058 Erlangen, Germany}
\vspace{-10mm}
\pacs{42.50.-p, 42.50.Lc, 42.50.Ar, 42.65.-k}

\begin{abstract}
Applying a multiphoton-subtraction technique to two-color macroscopic squeezed vacuum state of light generated via high-gain parametric down conversion we conditionally prepare a new state of  light: bright multi-mode low-noise twin beams. The obtained results demonstrate up to 8-fold suppression of noise in each beam while preserving and even moderately improving the nonclassical photon number correlations between the beams. The prepared low-noise macroscopic state, containing up to $2000$ photons per mode, is not among the states achievable through nonlinear optical processes. The proposed technique substantially improves the usefulness of twin beams for quantum technologies.  
\end{abstract}

\maketitle
\narrowtext
\noindent Quantum nonclassical states of light are important for fundamental tests of quantum theory and, as well, they are basic resources for quantum technology. Perfect correlations of photon numbers in twin beams generated in an optical parametrical amplifier (OPA) with no signal or idler input~\cite{Christine2011, Perina2005} make these states applicable for quantum metrology~\cite{Brida2006}, quantum imaging~\cite{BridaNP2010}, and quantum information~\cite{MadsenNC2010}. Fluctuations of the photon-number difference in twin beams are completely suppressed at the expense of the absolutely uncertain relative phase. Theoretically, the nonclassicality witnessed by that noise suppression is conserved for any parametric gain leading to the unlimited brightness of nonclassical radiation.
Unfortunately, thermal photon-number distribution of each beam~\cite{Bondany2004} makes these states very noisy and less suitable for super-sensitivity experiments~\cite{Mimih}, for the preparation of higher-order Fock states, and for certain quantum communication applications~\cite{Vlad2005}. Therefore one should develop methods which will allow to generate a new type of states, namely low-noise bright squeezed vacuum twin beams. 

In this Letter, we suggest and implement the method for the conditional preparation of macroscopic nonclassical twin-beam states with suppressed noise. The method is based on a macroscopic extension of photon subtraction technique~\cite{Ulrik2008, Wenger2004, Parigi2007, Sasaki2010, Zhai2013}. We show that it reduces the noise in the photon-number distributions of the individual beams characterized by mean-to-deviation ratio, while preserving strong photon-number correlation between them.
Moreover, the noise is reduced independently of the mode structure of the twin beams without the increase of the number of the modes. Thus, we demonstrate the preparation of low-noise bright twin-beam states of light which cannot be deterministically obtained by processes of parametric down conversion or four-wave mixing.

\begin{figure} [h]
\centerline{\includegraphics[width=6cm]{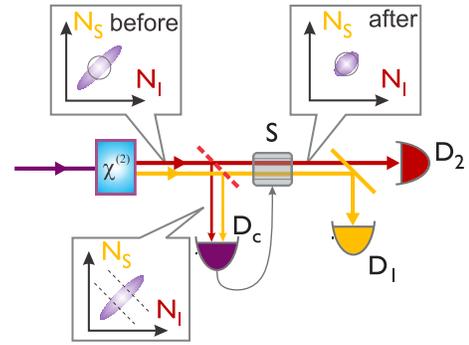}}
\caption{(Color online) A principle scheme of the experiment. Joint photon-number distribution of the signal and idler beams generated via parametric down conversion at the output of the nonlinear crystal $\chi^{(2)}$ has an elliptic shape. Circles display the two-mode shot noise limit. A weak measurement of the photon numbers in both beams is implemented by a low reflecting beam splitter (a dashed red line) and a detector $D_{c}$. The dashed lines denote a selection interval for a sum of the photon numbers. Detection of the particular number of photons in $D_c$ opens the shutter S and leads to the reduction of the noise in each of the beams (shown as an ellipse with a major axis reduced) which are detected by two independent detectors $D_1$ and $D_2$.}
\label{idea}
\end{figure}
{\em Method.-} To prepare low-noise twin beams we start from the pure state
\begin{equation}\label{ini}
|\Psi\rangle=\frac{1}{\sqrt{1+N_m}}\sum_{n=0}^{\infty}\lambda^{\frac{n}{2}}
|n\rangle_S|n\rangle_I
\end{equation}
of the single-mode signal (S) and idler (I) beams, where $N_m$ is the mean number of photons present in the beam and $\lambda=\frac{N_m}{N_m+1}$. The statistics of the beams are thermal and the state exhibits ideal photon-number correlation, i.e. the number of photons in S is exactly the same as the number of photons in I.

In the multi-mode case with $M$ matched modes, the state can be
effectively described by $|\Psi^{(M)}\rangle=\sum_{n=0}^{\infty}\sqrt{p_n^{(M)}}|n^{\otimes}\rangle_S|n^{\otimes}\rangle_I$
with $|n^{\otimes}\rangle=\delta\left(n-\sum_{l=1}^{M}n_l\right)|n_l\rangle$, where
\begin{equation}
p_n^{(M)}(N_m)=\frac{(n+M-1)!}{n!(M-1)!\left(N_m+1\right)^M\left(N_m^{-1}+1\right)^n}
\end{equation}
\noindent and the perfect photon-number correlation is preserved. The noise in individual beams is suppressed by increasing the number of modes, but it reduces coherence properties of the generated state. In this work we look for a method which is able to decrease the noise in the twin beams without increasing the number of modes. The principle idea of the experiment is shown in Fig.~\ref{idea}, together with a description.

In the quantum description, subtraction of N photons from S and I beams transforms the single-mode state (\ref{ini}) to the state
\begin{equation}\label{onemodesub}
|\Psi'\rangle=\frac{\sum_{n=0}^{\infty}\frac{(n+N)!}{n!N!}
\lambda^{\frac{n}{2}}
|n\rangle_S|n\rangle_I}{\sqrt{_2F_1\left[1+N,1+N,1,\lambda\right]}}
\end{equation}
where $_2F_1$ is the hypergeometric function. The photon correlation is kept, but
interestingly, with the increase of the number $N$ of subtracted photons the beams gradually change their statistics so that the  peak of photon-number distribution moves away from the origin. Consequently, the brightness of the individual
beams surprisingly increases linearly with $N$. The beams therefore become simultaneously more
bright and less noisy. It can be efficiently described by a {\em mean-to-deviation
ratio} (MDR), defined as $MDR=\frac{\langle n\rangle}{\sqrt{\langle(\Delta n)^2\rangle}}$, which {\em universally} increases with the number $N$ of subtracted photons for any $N_m$, being approximately proportional to $\sqrt{N}$ at large $N$ (it can also have the meaning of signal-to-noise ratio in particular applications such as quantum imaging).  The reason for using MDR to characterize the statistics is given in the Supplemental Material~\cite{SM}. In the multimode regime, the total mean number of photons $\langle n\rangle=MN_m$ and $\mbox{MDR}=\sqrt{M}\frac{N_m}{1+N_m}$ can both grow but only by increasing the number of modes $M$, which reduces the spectral mode purity. The noise reduction obtained by our photon subtraction preserves the mode structure of the initial radiation spectrum. 
 Moreover, we can conditionally prepare novel low-noise macroscopic nonclassical states which are
not compatible with any combination of nonlinear
optical processes.

\begin{figure} [htb]
\centerline{\includegraphics[width=9cm]{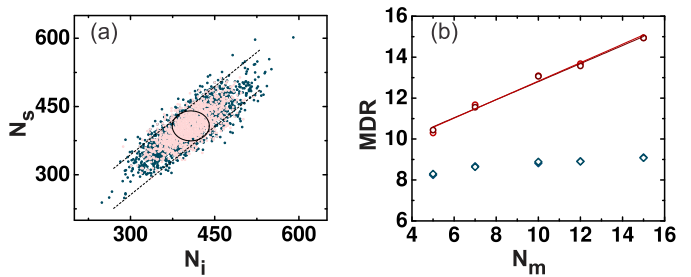}}
\caption{(Color online) (a): photon-distribution of the initial modeled state (blue points) with $N_m=7$, $M=91$ matched and $K=9$ unmatched modes, detection efficiency $80\%$, $10\%$ of tapping, and the conditionally prepared state (red circles). The solid black circle shows the shot-noise width. Dashed lines indicate the standard deviation region of the photon-number difference. The state demonstrates just $12\%$ of twin-beam squeezing due to high the influence of unmatched modes. (b): MDR of the signal and idler beams of the numerically modeled multimode state versus the mean photon number per mode $N_m$. Results for the S and I beams before the subtraction coincide and are shown by diamonds. Circles with linear fits clearly demonstrate the increase of the MDR for the twin beams after the photon subtraction.} 
\label{theory}
\end{figure}

{\em Numerical simulation for the multi-mode case.-} We numerically analyze the method of photon subtraction in the general case of multi-mode states, taking into account realistic conditions. The numerical model was built using random number generators preparing bipartite initially perfectly correlated thermal photon-number distributions governed by the
twin-beam probability distribution $p(n)=N_m^n/(N_m+1)^{n+1}$ in each of the statistically
independent modes. The conditioning (photon-subtraction) was applied as splitting of both signal and idler beams on an
unbalanced beamsplitter (tapping) and subsequent joint mode-nondiscriminating photon-number measurement
on the tapping detector. This can be considered as a weak measurement of the photon number in both beams. Then the remaining signal and idler modes were detected by other mode-nondiscriminating
photon-counters, and only the outcomes, satisfying a certain condition for the photon number registered by the tapping detector, were kept. The condition is defined as $n_1<n<n_2$, where $n$ is the number of photons registered at the control detector; $n_1$ and $n_2$ define the condition bounds. The condition is symmetrical with respect to the mean photon number at the tapping detector, thus it is not changing the mean photon number of the state. The width $n_2-n_1$ of the condition is set to half of the standard deviation of the photon number at the control detector. The realistic experimental conditions were simulated as finite detection efficiency (modeled as mode-wise coupling to vacuum prior to photon counting) and mode mismatch~\cite{Ivantwocolor}, which reduces the photon-number correlation of the detected state. The signal-idler distribution for a state with $N_m=7$, $M=91$ matched and $K=9$ unmatched modes generated with $2\times 10^4$ points is given in Fig. \ref{theory}(a) prior to conditioning (blue points) and after the conditioning (red circles). Joint distributions look similar to the distribution obtained for the Sub-Poissonian light in~\cite{Leuchs98}, but with $90^{\circ}$ rotated orientation demonstrating the noise suppression in the photon-number difference.

The results of photon subtraction in terms of MDR for both signal and idler beams versus $N_m$ are given in Fig. \ref{theory} (b), where the ratio is given before (lower points) and after the conditioning  applied (upper points and overlapping linear fits). It is clearly visible that subtraction reduces the noise in the individual beams for the multimode states even in the case of a realistic measurement with inefficient detectors and mode mismatch. Thus, the single-mode effect survives in the multi-mode case, resulting in a macroscopic effect.

Since conditioning does not change the mean photon number, the effect of MDR increase is clearly governed by the effective suppression of the photon-number variance in the beam.

We also verify that the correlation properties of twin-beam states are not degraded by the described method. In order to characterize the signal-idler correlations we use the standard approach based on the noise reduction factor (NRF). NRF is defined as the variance of the photon-number difference in the signal ($n_s$) and idler ($n_i$) beams normalized to the mean total number of photons, $NRF \equiv \hbox{Var} (n_i-n_s)/\langle n_i+n_s \rangle$. Considering each mode of the PDC radiation to be thermal and assuming that each detected beam contains $M$ matched modes and $K$ unmatched ones, we can evaluate the $NRF_{meas}$ that is observed in the experiment as
\begin{equation}
NRF_{meas}=1-\frac{M}{M+K}\eta+\frac{K}{M+K}\eta N_m.
\label{NRF}
\end{equation}
\noindent Here $\eta$ is the total detection efficiency.

Calculations show that NRF is not increased by applying the condition. Thus, the suggested method of beam noise reduction by conditioning is indeed efficient in the case of multimode states and realistic measurements. Further we confirm the proposed method experimentally for bright multimode twin beams, whose statistics is hard to simulate numerically due to the high occupation numbers in the modes. Experiment is thus a crucial step allowing to verify the effect in the macroscopic regime.

{\em Experimental verification.-} To verify our method in the experiment we have built a setup which is described in detail in the Supplemental Material~\cite{SM}. Bright two-color twin beams of PDC were generated in an optical parametric amplifier pumped by  the third harmonic of a high-power pulsed Nd:YAG laser and were detected by two separate detectors. Photon-number occupation of the PDC was changed with the intensity of the pump. Multi-photon subtraction was performed by detecting a small fraction of the signal and idler beams by the detector $D_c$. In the measurement, varying the pump power, the total number of subtracted photons per pulse was in the range of $30000<n<180000$ on the average.
\begin{figure} [h]
\centerline{\includegraphics[width=8cm]{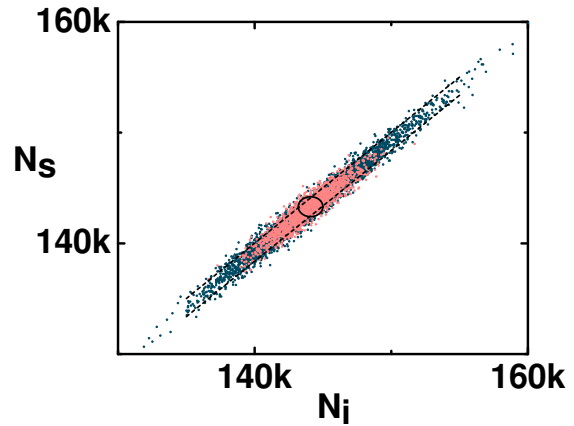}}
\caption{(Color online) Experimentally measured joint distribution for the twin beams before (blue points) and after (red points) multi-photon subtraction with $N_m=390$, $M+K=1000$. The solid black line (circle) shows the noise width determined by the two mode photon shot noise and the electronic noise of the detectors (530 electrons/pulse). Dashed lines indicate the standard deviation region of the photon-number difference for the state demonstrating $50\%$ of twin-beam squeezing. 
The plot explicitly shows the efficiency of the multi-photon subtraction, demonstrating a reduction the noise in each of the beams, while retaining the nonclassical correlations.}
\label{Jointexp}
\end{figure}
To simplify the experiment we have performed a procedure equivalent to the standard feedforward technique~\cite{Fabre2003}. Depending on the result of the measurement in the detector $D_c$ the photon numbers measured in the signal and idler beams were accepted or discarded for the calculation of the $MDR$ and $NRF$. The signals in the detectors $\mathrm{D_1, D_2}$ were used only under condition that the signal per pulse $\mathrm{S_D}$ in the detector $\mathrm{D_c}$ took a value within the range of $\sigma/15$ around the mean signal $\langle\mathrm{S_D}\rangle$. Here $\sigma$ is the standard deviation of $S_D$.  Fig. \ref{Jointexp} shows experimentally measured joint distributions for the conditionally prepared state (red circles) together with the distribution for the initial state (blue points). This result qualitatively verifies our numerical expectation shown in Fig. \ref{theory} (a).
\begin{figure} [h]
\centerline{\includegraphics[width=9.0cm]{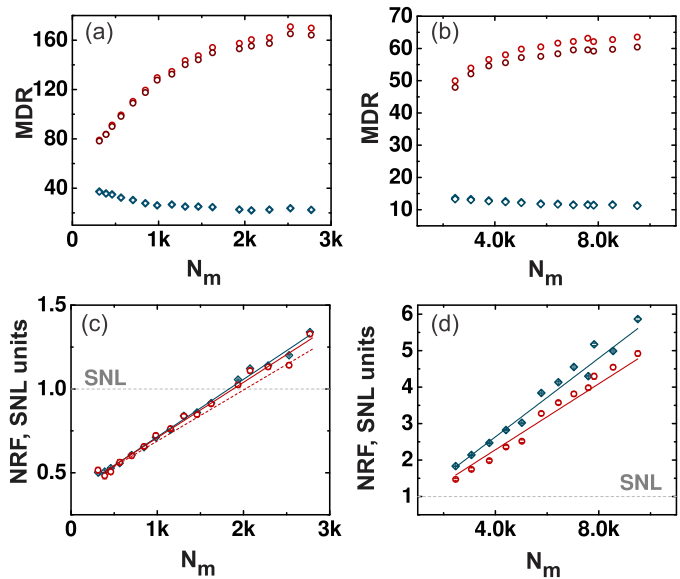}}
\caption{(Color online) Mean-deviation ratio $\mathrm{MDR}$ and noise reduction factor $\mathrm{NRF}$ measured versus the number of photons per mode $N_m$. The number of modes in the figures (a, c) was approximately five time as large as in (b, d). Blue diamonds and red (brown) circles show the results before and after conditioning, respectively. Straight red and blue lines are linear fits. A dashed line shows a result of conditioning in the range of $(0.93\cdot \langle S_D\rangle \pm \sigma/30)$.}
\label{fig:2}
\end{figure}
Our method was applied to the PDC radiation containing different numbers of modes, which was realized by using two different sets of apertures $\mathrm{A_1,A_2}$ placed in front of the detectors $D_1$ and $D_2$. In the first measurement the diameters of the irises were $\mathrm{4\;mm}$ and $\mathrm{5.14\;mm}$. Fig.~\ref{fig:2} (a) presents MDR measured as a function of $\mathrm{N_m}$ before (blue diamonds) and after conditioning (red and brown circles for the signal and idler beams, respectively). The observed reduction of the unconditional MDR with the growth of $N_m$ is caused by the reduction of the number of the detected modes in the range of $\mathrm{1000<M+K<500}$ with the increase of the parametric gain~\cite{g2}. One can see that the procedure of conditioning significantly (up to 8 times) reduces the noise of the photon-number distributions in the signal and idler beams. As expected, different noise suppression in the signal and idler channels is observed due to the fact that the tapping signal contains different numbers of modes for each beam. Figure \ref{fig:2} (b) represent analogous data in the case where the aperture sizes were reduced to $\mathrm{2\;mm}$ and $\mathrm{2.56\;mm}$. In order to work within the dynamic range of our detectors, we increased the intensity of the pump. One observes that for the state with reduced number of modes ($\mathrm{180<M+K<130}$) the increase of the MDR after the conditioning (shown in red) with respect to the MDR before the conditioning (shown in blue) is also efficient as in the multi-mode case.

To prove that the conditioning preserves and even improves the nonclassical photon-number correlations, we show linear dependencies of the NRF before (blue diamonds) and after (red circles) the conditioning in Fig. \ref{fig:2}(c,d) measured with the same sets of the apertures as MDR. As shown in Fig. \ref{fig:2}(c), for larger apertures we have obtained almost overlapping straight solid lines plotted according to (\ref{NRF}) in the assumption that $m\gg k$. For the quantum efficiency we obtained the fitting values $\eta_b = 0.63 \pm 0.01$ before and after $\eta_a= 0.62\pm 0.01$, respectively, in a good agreement with the estimated detection efficiency ($\eta_{det} \cdot \eta_{PBS_1} \cdot \eta_{optics}=0.82\cdot0.88\cdot0.9=0.63$). Moreover, as shown in the Supplemental Material~\cite{SM}, the independent measurement of the normalized variance of the photon number sum provides the same $\eta=0.64\pm 0.02$. Insignificant decrease of the slope after conditioning with respect to the slope obtained before the conditioning demonstrates reduction of the impact of the unmatched modes. It is important to mention here that considering the signal in the tapping channel in the range of $(0.93\cdot \langle S_D\rangle \pm \sigma/30)$  one can obtain a new linear dependence (red dashed line) with the slope reduced by $10\%$. In this case the applied method even increases the brightness of the twin-beam state with the measured sub-Poissonian photon-number correlation up to $2020$ photons per mode.

The measurement with the small apertures illustrates that the method is more efficient for the state with low number of modes. As shown in Fig.~\ref{fig:2}(d) the mode-number reduction leads to the increase of the unmatched modes influence: the slope of the linear fit before the conditioning is much steeper. Due to the high brightness of the state we did not experimentally observe the reduction of the noise below the shot noise level. However, in this case the method of the noise suppression was more efficient: the value of the slope after conditioning is reduced by 20\%.

{\em Conclusion.-} In conclusion, we have experimentally prepared a new type of twin-beam squeezed vacuum states. Our results show that photon macroscopic version of photon subtraction technique allows one to suppress the fluctuations in each beam while preserving and even improving the observable degree of two-mode squeezing. Sub-shot-noise photon-number correlations were observed for the low-noise twin beams containing up to 2000 photons per mode. This approach can be stimulating for spin-squeezing experiments with macroscopic number of atoms~\cite{Giacobino, Mitchell}.  We believe that the suggested method is a useful step towards generation of macroscopic multi-mode states of light which demonstrate essentially quantum properties beyond standard nonlinear quantum optics.

We acknowledge partial financial support of the EU FP7 under grant agreement No. 308803 (project BRISQ2), V.C.U. acknowledges the project 13-27533J of GA\v CR.

\end{document}


\title{Supplemental Material: Low-noise bright twin beams}
\author{Timur Sh.~Iskhakov}
\affiliation{Max Planck Institute for the Science of Light,
G\"unther-Scharowsky-Stra\ss{}e 1/Bau 24, 91058 Erlangen, Germany}
\author{Vladyslav~C.~Usenko}
\affiliation{Department of Optics, Palack\'y University, 17. Listopadu 50, 772 00 Olomouc, Czech Republic}
\affiliation{Bogolyubov Institute for Theoretical Physics, Metrolohichna str. 14-b, 03680 Kiev, Ukraine}
\author{Radim Filip}
\affiliation{Department of Optics, Palack\'y University, 17. Listopadu 50, 772 00 Olomouc, Czech Republic}
\author{Maria~V.~Chekhova}
\affiliation{Max Planck Institute for the Science of Light, G\"unther-Scharowsky-Stra\ss{}e 1/Bau 24, 91058
Erlangen, Germany} \affiliation{Department of Physics, M.V.Lomonosov Moscow State University, \\ Leninskie Gory,
119991 Moscow, Russia}\affiliation{University of Erlangen-N\"urnberg, Staudtstrasse 7/B2, 91058 Erlangen, Germany}
\author{Gerd Leuchs}
\affiliation{Max Planck Institute for the Science of Light, G\"unther-Scharowsky-Stra\ss{}e 1/Bau 24, 91058
Erlangen, Germany} \affiliation{University of Erlangen-N\"urnberg, Staudtstrasse 7/B2, 91058 Erlangen, Germany}
\vspace{-10mm}
\pacs{42.50.-p, 42.50.Lc, 42.50.Ar, 42.65.-k}

\begin{abstract}
In this text we theoretically compare different approaches for the state characterization based on the second-order intensity correlation function, the Fano-type factor, and the mean-to-deviation ratio measurement. Then we experimentally estimate the quantum efficiency of the detection based on the Fano factor measurement.

\end{abstract}
\vspace{5mm}
\maketitle \narrowtext
\section{Statistics characterization.}

\textit{Perfect subtraction.--} We first derive the result of a perfect N-photon subtraction applied to a twin-beam state $|\Psi\rangle$ (see the main text for the definition) as
\begin{equation}
(a_Sa_I)^N|\Psi\rangle=\frac{1}{\sqrt{1+N_m}}\sum_{n=0}^{\infty}\frac{(n+N)!}{N!}\lambda^{\frac{n+N}{2}}|n\rangle_S|n\rangle_I,
\end{equation}
where $a_{S,I}$ are the annihilation operators of the signal and idler beams respectively and $\lambda=\frac{N_m}{N_m+1}$.

The conditioning on the $N$ perfectly subtracted photons leads to the conditionally prepared normalized state
\begin{equation}\label{onemodesub}
|\Psi'\rangle=\frac{\sum_{n=0}^{\infty}\frac{(n+N)!}{n!N!}
\lambda^{\frac{n}{2}}
|n\rangle_S|n\rangle_I}{\sqrt{_2F_1\left[1+N,1+N,1,\lambda\right]}}
\end{equation}
where $_2F_1$ is the hypergeometric function.

Statistics of such twin beam state is defined by the variance and the mean value of the number of photons, but the statistical measures can be different. In particular, $g^{(2)}$ and the Fano $F$ factor are typically used in literature. We study the behavior of these parameters for the perfectly conditioned state.

\textsl{$g^{(2)}$ measure.--} $g^{(2)}$ is defined as the second-order intensity correlation function which can be expressed through the mean photon numbers in either signal or idler beams as $g^{(2)}\equiv(\langle n^2 \rangle-\langle n \rangle)/\langle n \rangle^2$.
The $g^{(2)}$ function always approaches unity as the number $N$ of subtracted photons increases. However,  $g^{(2)}$ is also sensitive to the number of modes of the twin beams. $g^{(2)}(0)=1+\frac{1}{M}$ can be reduced by increasing the number of thermal single-mode twin beams so that the resulting multi-mode state will have  $g^{(2)}$ close to $1$. The single-mode statistical contribution in $g^{(2)}$ is therefore not visible.

One should mention that for $N_m\ll 1$, the sub-Poissonian statistics with the second-order intensity correlation function
$g^{(2)}$ approaching $0.5$ can be obtained for some number of
subtracted photons, however, it continuously degrades as $N$ increases
further. For $N_m>1/3$ the conditional sub-Poisson statistics
completely vanishes, it is therefore not relevant for the bright twin
beams.

\textsl{Fano-type factor.--} Alternatively, the statistics of the beams can be characterized by the Fano-type factor $F$, being the normalized variance of the photon-number sum  $F\equiv \hbox{Var}(n_i+n_s)/\langle n_i+n_s \rangle$, which, in our case (assuming detection efficiency $\eta$), reads
\begin{equation}
F=2\eta N_m+1.
\label{F}
\end{equation}
It is reduced for the large number of subtracted photons $N$, but only weakly for a large $N_m\gg 1$.
In addition, $F$ can approach unity simply by attenuation. Thus is does not clearly characterize the change of the statistics of the beams due to subtraction and conditioning in all cases.

\textsl{Mean to deviation ratio.--} The reciprocal of the coefficient of variation, the \textsl{mean to deviation ratio} (MDR), describes the change of the standard deviation compared to the mean value of the data. MDR is actively used in the tasks such as imaging and has the meaning of the signal-to-noise ratio \cite{imaging}.

MDR can be defined through the photon number in either signal or idlear beam and can be approximately given as
\begin{equation}
\mbox{MDR}=\frac{\langle n\rangle}{\sqrt{\langle(\Delta n)^2\rangle}}
\approx \left\{
  \begin{array}{ll}
    \sqrt{2\lambda N -1} & \quad \lambda\ll 1\, \mbox{and large}\, N,\\
    \sqrt{2\lambda N +1} & \quad \lambda\rightarrow 1\, \mbox{and large}\, N.
  \end{array}\right.
\label{MDR}
\end{equation}
The analysis shows that MDR is sensitive to photon subtraction and conditioning both in single-mode and multi-mode cases for high and low intensities of the beams (see main text). Thus it is used in the paper to confirm the effect of photon subtraction and conditioning in both the single-mode and multi-mode case for any intensity of the beams.

\section{Experimental setup.}
\begin{figure} [h]
\centerline{\includegraphics[width=8cm]{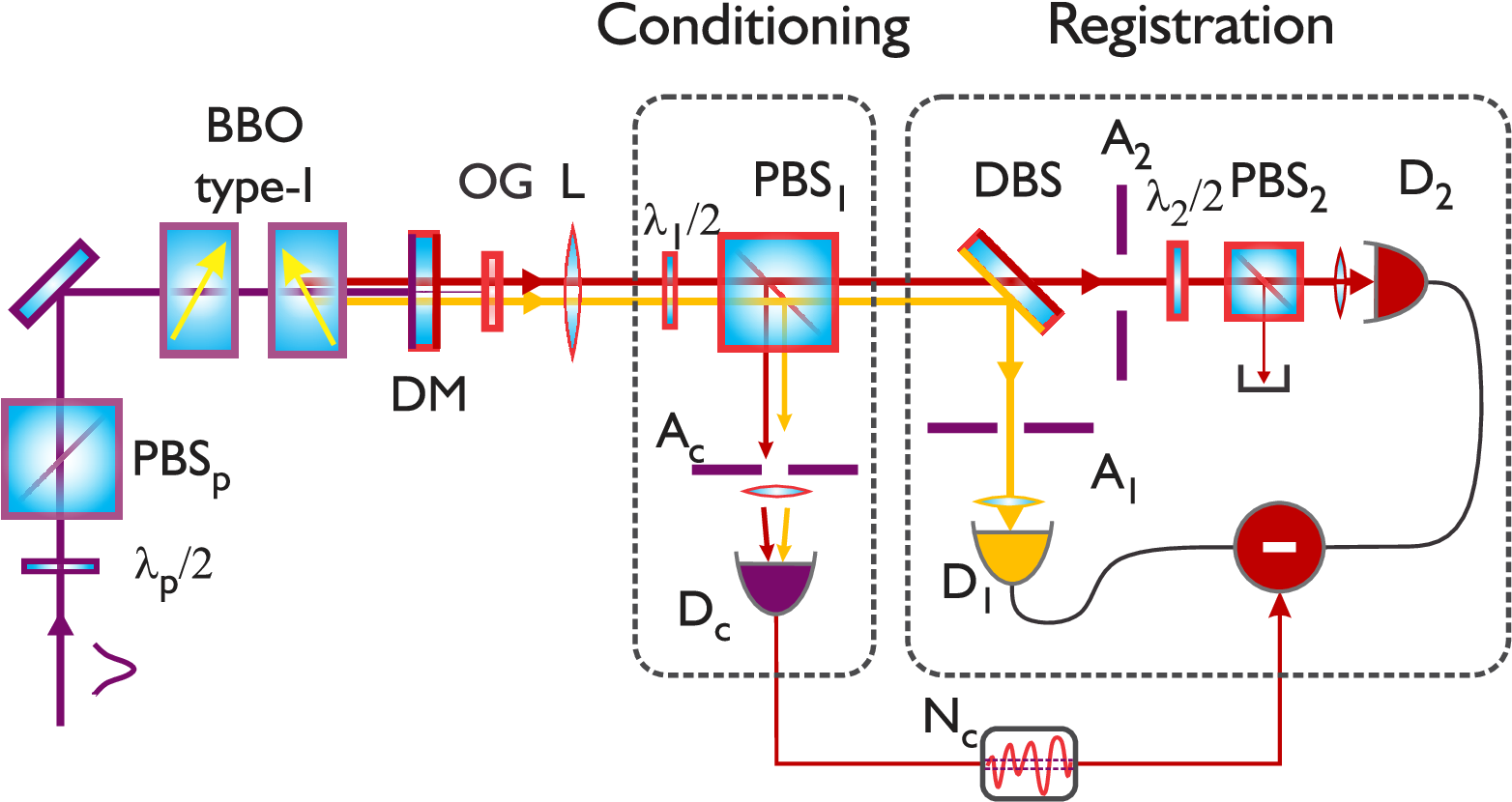}}
\caption{(Color online) The experimental setup.}
\label{fig:1}
\end{figure}

To verify our method in the experiment we have built a setup shown in Figure \ref{fig:1}. As a pump we have used third harmonic of the pulsed Nd:YAG laser at the wavelength 355 nm, with the repetition rate 1 kHz, pulse duration 18 ps, and energy per pulse up to 0.2 mJ. The laser beam waist was $\mathrm{210~\mu m}$. Twin beams at the central wavelengths of 635 nm (signal) and 805 nm (idler), respectively, were generated by the OPA based on two 3 mm thick type-I BBO crystals. The optical axes of the crystals were placed in the horizontal plane as shown in yellow in order to reduce the Poynting vector walk-off. The pump intensity was changed by rotation of a half-wave plate ($\lambda_p/2$) placed before the polarization beamsplitter ($\mathrm{PBS_p}$). After the crystals the pump beam was eliminated by a dichroic mirror $\mathrm{DM}$ with reflectance greater than 99.5 at 355 nm and transmission greater than 98\% at $\mathrm{\lambda_1=635\;nm}$ and $\mathrm{\lambda_2=805\;nm}$. The rest of the UV radiation was blocked by the long-pass filter $\mathrm{OG580}$. In the registration part of the setup signal and idler beams were separated in wavelength by means of a dichroic beamsplitter (DBS) and spatially restricted by two iris apertures ($\mathrm{A_{1,2}}$) placed in the focal plane of the lens L ($\mathrm{f=20\;cm}$). The diameters of the irises satisfied the condition $D_1/D_2=\lambda_1/\lambda_2$~\cite{Brambila2004}, providing equal numbers of the detected modes in both channels. After the irises all the radiation was focused by two 5 cm collecting lenses onto the Si PIN diode-based detectors Hamamatsu $\mathrm{S3072}$ with quantum efficiency of 82\% at 635 nm ($\mathrm{D_1}$) and Hamamatsu $\mathrm{S3883}$ with the efficiency of 89\% at 805 nm ($\mathrm{D_2}$). Detection efficiencies were optically balanced by means of a zero-order half-wave plate at 805 nm ($\lambda_2/2$) and a polarization beamsplitter ($\mathrm{PBS_2}$) placed in front of detector $\mathrm{D_2}$. The signals from the detectors were analyzed using 60 MS/s, 12-Bit, 8-channel digitizer NI PXI-5105 by National Instruments through integrating the electronic pulses over time. The data was stored in a computer for further processing.

The heart of the experiment is the conditioning part. We reflected 12\% of each beam using a combination of an achromatic half-wave plate $\mathrm{\lambda_1/2}$ and a broadband polarization beamsplitter $\mathrm{PBS_1}$. Then, the angular spectrums of both wavelengths were restricted by the iris $\mathrm{A_c}$ with the same size as $\mathrm{A_{1}}$ placed in the focal plane of the lens L. The transmitted radiation was focused by 5 cm collecting lens onto the tapping detector $\mathrm{D_c}$ similar to the detector $\mathrm{D_2}$. The signal from the detector was also sent to the digitizer and stored in the computer. In the experiment the pump power was varied and for each value of the pump power 300 000 pulses were accumulated.
The parametric gain was evaluated from the nonlinear dependence of the PDC signal $S$ on the pump power. For this measurement we have installed a pinhole of diameter $800\mu m$ in the focal plane of the lens L and a bandpass filter centered at $\lambda=635nm$ (10nm bandwidth) in front of the detector $D_1$. The obtained dependence was fit according to the formula
\begin{equation}
\langle S\rangle=A \cdot \sinh^2(\sqrt{B\cdot P}).
\label{gain}
\end{equation}
\noindent The brackets $\langle ...\rangle$ denote the average over 10000 pulses, $A$ and $B$ are the fitting parameters, $P$ is the pump power. The parametric gain $\mathrm{G=\sqrt{B\cdot P}}$ was varied in the range of $\mathrm{3.6 <G <5.3}$ by changing the pump power in the range of $\mathrm{12\;mW<P<28\;mW}$.
\section{Quantum efficiency estimation.}
\begin{figure}
[h]\centerline{\includegraphics[width=9.0cm]{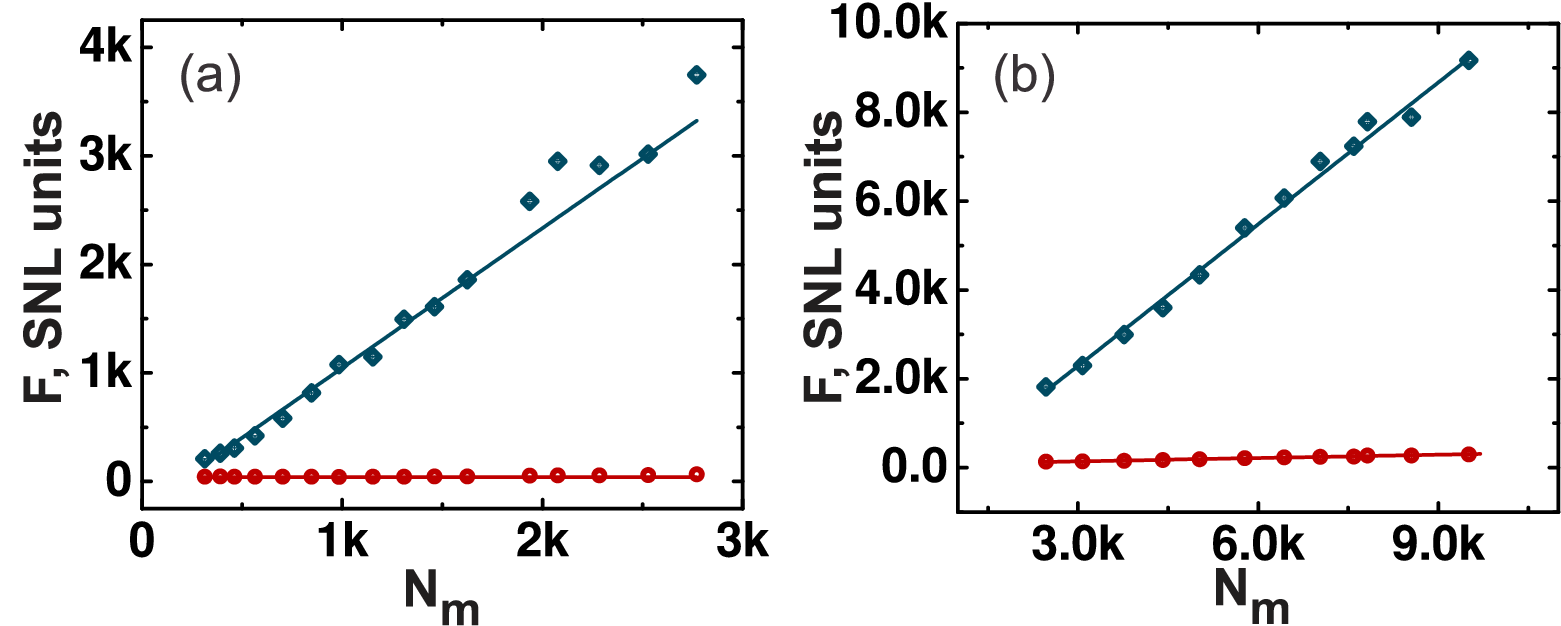}}
\caption{(Color online) Fano-type factor versus the number of photons per mode measured with big (a) and small apertures (b). Blue diamonds and red circles show the results before and after conditioning, respectively. Straight red and blue lines are linear fits.}
\label{fig:2}
\end{figure}

It is known that the quantum efficiency of a photodiode can be determined trough the noise reduction factor (NRF, see the main text for the definition) measurement~\cite{Ivano}. According to formula (\ref{F}) 
the quantum efficiency can also be estimated from the slope of $F$ vs $N_m$ dependence. In order to implement this method we have calculated the normalized variance of the photon-number sum as a function of the number of photons per mode (Fig.\ref{fig:2} (a)). The F parameter was evaluated using the same data array that was collected for the $NRF$ and $MDR$ presented in Fig.3 (a,b) in the main text. The values of the Fano factor before and after conditioning are shown by blue diamonds and red circles, respectively. Solid lines are the fits, plotted according to (\ref{F}). One can see that the blue points perfectly fit the linear function. The slope of this line was found to be $1.28\pm0.05$. The result is in remarkably good agreement with the theoretical expectation of $2\cdot\eta = 1.26$, given by the losses in the optical channels, and the value obtained independently in the NRF measurement through Eq. (4) in the main paper, with the experimentally measured values of $\eta_b = 0.63 \pm 0.01$ and $\eta_a= 0.62\pm 0.01$. 
After conditioning 
the slope of the line was reduced to $0.007\pm0.001$ demonstrating 180-fold reduction of the noise influence.
Similar measurements were performed with the small apertures. The results are shown in (Fig.\ref{fig:2} (b)). The obtained slope of $F$ on $N_m$ took the value of $2\eta=1.06\pm0.02$, demonstrating the decrease of the detection efficiency caused by a strong spatial filtering.  After conditioning 
the slope was found to be $0.025\pm0.001$, which is significantly smaller then the initial value.

